\documentclass[proof]{pasj00}
\draft
\begin{document}
\SetRunningHead{T.Yoda et al.}{The AMANOGAWA-2SB Galactic Plane Survey I}
\Received{2010/03/18}
\Accepted{2010/07/23}
\title{The AMANOGAWA-2SB Galactic Plane Survey I:\\ Data on the Galactic Equator}
\author{Takahiro \textsc{Yoda}, Toshihiro \textsc{Handa}, Kotaro \textsc{Kohno}}
\affil{Institute of Astronomy, The University of Tokyo, 2-21-1 Osawa, Mitaka, Tokyo 181-0015}
\email{yoda@ioa.s.u-tokyo.ac.jp}
\author{Taku \textsc{Nakajima}\thanks{Present address: Nobeyama Radio Observatory, National Astronomical Observatory of Japan, 462-2 Nobeyama, Minamimaki, Minamisaku, Nagano 384-1305}, Masahiro \textsc{Kaiden}, Yoshinori \textsc{Yonekura}\thanks{Present address: Center for Astronomy, Ibaraki University, 2-1-1 Bunkyo, Mito, Ibaraki 310-8512}, Hideo \textsc{Ogawa}}
\affil{Department of Physical Science, Graduate School of Science, Osaka Prefecture University, 1-1 Gakuen-cho, Naka-ku, Sakai, Osaka 599-8531}
\author{Jun-ichi \textsc{Morino}}
\affil{National Astronomical Observatory of Japan, 2-21-1 Osawa, Mitaka, Tokyo 181-8588}
\and
\author{Kazuhito \textsc{Dobashi}}
\affil{Department of Astronomy and Earth Sciences, Tokyo Gakugei University, 4-1-1 Nukuikita-machi, Koganei, Tokyo 184-8501}

%


\maketitle

\begin{abstract}

Using a waveguide-type sideband-separating receiver (2SB receiver) on the Tokyo-NRO 60-cm telescope (renamed the AMANOGAWA telescope), we carried out simultaneous observations in the $^{12}$CO($J=2-1$) and $^{13}$CO($J=2-1$) lines over the Galactic plane $l=10^\circ-245^\circ$ along $b=0^\circ$ with a $\timeform{3'.75}$ grid. Using the $^{12}$CO($J=1-0$) data of Dame et al. (2001), who used a beam size almost the same as ours, we show $^{12}$CO($J=2-1$)/$^{12}$CO($J=1-0$) and $^{13}$CO($J=2-1$)/$^{12}$CO($J=2-1$) intensity ratios on the $l-v$ map and the intensity correlations among the $^{12}$CO($J=2-1$), $^{13}$CO($J=2-1$), and $^{12}$CO($J=1-0$) lines. As a result, a linear correlation between $^{12}$CO($J=1-0$) and $^{12}$CO($J=2-1$) and a curve correlation between $^{12}$CO($J=2-1$) and $^{13}$CO($J=2-1$), as produced by most of the data, are found. We investigate these correlations with simple radiative transfer equations to ascertain a number of restrictions on the physical quantities of molecular gas on a galactic scale.

\end{abstract}

\section{Introduction}\label{sec:intro}

In order to understand the dynamics and evolution of galaxies, it is important to study the distribution and the physical condition of interstellar gases. In particular, molecular gas is important because it forms stars, which occupy about 90\% of the visible mass of their host galaxies. The Schmidt law gives the relation between the star formation rate and the mass surface density of the gas (Schmidt, 1958). As a result of observational investigation, its power index has been determined to be about 1.4 (Kenicutt, 1998).

In one empirical scenario for the formation of massive stars, stars are formed in a molecular core whose density is remarkably high, at about $10^7$cm$^{-3}$ (e.g. Gracia-Carpio et al. 2008; Muraoka et al., 2009). On the other hand, the fact that the Schmidt law remains valid for a giant molecular cloud (GMC) scale of $\sim10^2$ pc suggests a relationship between the molecular cores and the gas properties averaged over a GMC. In order to investigate this relationship, it is necessary to obtain high resolution data on a galactic scale. The Milky Way Galaxy (MWG) is the best target source for this purpose, because it can be observed with the highest spatial resolution and S/N ratio. We can elucidate the relationship between the physical condition of GMCs and the formation of stars in their environment through detailed observation. If the data cover the greater part of the MWG and does not omit the less dense gases, we can compare them with the data of external galaxies, whose data are smoothed over a large area with a telescope's beam.

Dame et al. (2001) carried out a survey of the entire MWG in $^{12}$CO($J=1-0$) and showed the distribution of molecular gases, including less dense gases. Observations coupled with this data make it possible for us to investigate the physical conditions of the molecular gases. If the observations are carried out in the line of another transition of CO or its isotope, we can minimize the uncertainty introduced by chemical abundance.

Sanders et al. (1986) and Jackson et al. (2006) carried out surveys that covered a large area of the MWG in $^{12}$CO($J=1-0$) and $^{13}$CO($J=1-0$), respectively. However, the coverage of their observations was limited to a part of the first Galactic quadrant and they did not observe the higher-$J$ transitions of the same molecular species.

Sakamoto et al. (1995) carried out a survey in $^{12}$CO($J=2-1$), whose coverage was also limited to $l=20^\circ-60^\circ$, using the same beam size as Cohen et al. (1986), and they found a decrease in the $^{12}$CO($J=2-1$)/$^{12}$CO($J=1-0$) intensity ratio with an increase in the Galacto-centric distance (Sakamoto et al., 1997). They could not, however, offer a detailed discussion of the 2-dimensional distribution due to the insufficient sensitivity of their receiver.

We upgraded the Tokyo-NRO 60-cm telescope used by Sakamoto et al. (1995) by installing a waveguide-type sideband-separating (2SB) SIS receiver. It then had a sensitivity that was 10 times better, and the two frequency bands could be observed simultaneously (Nakajima et al., 2007). Using this system, we began a new survey of both $^{12}$CO($J=2-1$) and $^{13}$CO($J=2-1$) over the Galactic plane.

As the first result of this survey, in this paper we present the data on the Galactic equator ($b=0^\circ$). In addition, using the survey data, we discuss the condition of the ``typical'' molecular gas in the MWG without a complicated molecular excitation model, as the first step in our investigation.

In Section 2, we describe the specifics of our observations. The data and comparisons of the three CO lines are given in Section 3. Using these results, in Section 4, we discuss the physical condition of the typical molecular gas in the Galactic disk. Finally, we present a summary in Section 5.

\bigskip

\section{Observations}
\subsection{{\rm Instrumentation}}\label{sec:instrument}

The telescope we used is the Tokyo-NRO 60-cm telescope (VST-1), which was specifically designed for rapid sky surveys on the 200 GHz band. It was built in 1990 and upgraded in 2006 by replacement of the receiver system. In this paper, we refer to the new telescope system as the AMANOGAWA telescope, which is an acronym for `A Miniature Antenna at NObeyama for GAlactic objects and the milky WAy'.

A 2SB receiver developed by Nakajima et al. (2007) was installed on our telescope. This is the first operated receiver of this type used for scientific observations on the 200 GHz band. It enables us to obtain two frequency band datasets on both sides of the local oscillator frequency; the upper sideband (USB) and the lower sideband (LSB). In addition, its waveguide-based design produces a lower noise temperature. The typical system noise temperature, including the atmosphere, is about 200 K at the zenith. Our spectra were obtained using two 2048-ch Acousto-Optical Spectrometers (AOS) developed by Kaiden et al. (2006). Their frequency coverage and resolution are 250 MHz and 250 kHz, respectively. These values correspond to a velocity coverage of 326 km s$^{-1}$ and a velocity resolution of 0.3 km s$^{-1}$ at 230 GHz, respectively. The basic characteristics and detailed specifications of the AMANOGAWA telescope are reported in Nakajima et al. (2007).

Although Nakajima et al. (2007) used the new moon, we found that it was not suitable for measurements of the main beam efficiency and the half power beam width (HPBW) of our telescope. The measured antenna temperature of the new moon changed from month to month, although the intensities of our calibration source were stable (see subsection 2.3). This can be attributed to the sensitive change of the beam averaged brightness by the moon phase and the position on the moon. The measured antenna temperature also was not stable for the full moon. Therefore, we measured the efficiency and HPBW using the sun in January 2009. In order to avoid saturation of the SIS mixer, we installed a wire grid in front of the receiver.

The sun is larger than the main beam of the telescope and the sidelobe level is sufficiently low. Measurement of the sidelobe level showed it to be less than -18 dB relative to the peak. In this case, the main beam efficiency $\eta_{\rm mb}$ can be estimated as
\begin{equation}
\eta_{\rm mb}=\frac{\int_{\rm mainbeam}P{\rm d}\Omega}{\int_{2\pi}P{\rm d}\Omega} \simeq \eta_{\rm sun} = \frac{\int_{\rm sun}P{\rm d}\Omega}{\int_{2\pi}P{\rm d}\Omega} = \frac{T_{\rm A,sun}^{*}}{T_{\rm sun}},
\label{eq:etamb}
\end{equation}
where $P$ is the beam pattern, $T_{\rm A,sun}^*$ is the measured antenna temperature of the sun, and $T_{\rm sun}$ is the intrinsic brightness temperature of the sun. The difference between $\eta_{\rm mb}$ and $\eta_{\rm sun}$ is estimated to be $\lesssim$1 \%. The $T_{\rm A,sun}^*$ is measured using attenuation rate $g$ of the wire grid as
\begin{equation}
T_{\rm A,sun}^{*} = \frac{g T_{\rm sun}-g T_{\rm sky}}{g T_{\rm hot}-g T_{\rm sky}}T_{\rm atm},
\label{eq:antenna}
\end{equation}
where $T_{\rm hot}$ is the temperature of the hot load and $T_{\rm sky}$ is the brightness temperature of blank sky. The temperature of the atmosphere $T_{\rm atm}$ is assumed to be equal to $T_{\rm hot}$. The estimated $\eta_{\rm mb}$ is free from the value of $g$. We carried out the measurement several times using the different values of $g$ that ranged from 0.25 to 0.75. Using a $T_{\rm sun}$ of 6256 K (USB) and 6299 K (LSB) (Linsky, 1973), we obtained a $\eta_{\rm mb}$ value of 0.725$\pm$0.029 for the USB and 0.767$\pm$0.038 for the LSB.

The HPBW was measured at $\timeform{8'.7}\pm\timeform{0'.4}$, a value that is quite close to that of the 1.2 m telescope used in the CfA CO($J=1-0$) survey conducted by Dame et al. (2001). The spatial resolution corresponds to 20 pc at 8 kpc from the sun.

The dataset we used in this work was obtained in 2007 and the measurement of the beam was revised in 2009. We maintained the same telescope optics, however, during this period. This means that the revised $\eta_{\rm mb}$ and HPBW are valid for all the data in this paper.

\bigskip
\subsection{{\rm Observations and Data Reduction}}\label{sec:obs}

The observations were carried out in the first season of the AMANOGAWA-2SB Galactic Plane Survey from January to April, 2007. We observed the $^{12}$CO($J=2-1$) and $^{13}$CO($J=2-1$) lines simultaneously. The sky coverage was $l=10^\circ-245^\circ$ along $b=0^\circ$, which is the entire range of longitude accessible from the observation site, except near the Galactic center. The grid spacing was $\timeform{3'.75}$ or 0.4 times that of HPBW. The doppler tracking velocity was fixed at $v_{\rm LSR}=0$ km s$^{-1}$ for each line. We obtained the data of the velocity range from -160 km s$^{-1}$ to +160 km s$^{-1}$ in all spectra.

All data were taken in a position-switching mode, in which off positions about 5$^\circ$ from the Galactic equator were selected from those given by Dame et al. (2001). After a 10-second integration period, the on position was switched to the off position. We summed up 6 to 8 sets to obtain the final integration data.

Due to the stable receiver system and offset optics, the baselines of the spectra are quite flat, so that only linear baselines were subtracted. After this procedure, Gaussian smoothing was applied to all the profile data in order to reduce the noise level. For the $^{12}$CO data, a Gaussian with 1.3 km s$^{-1}$ velocity width was applied, which is the same as the velocity resolution of Dame et al. (2001). The resultant noise level was $\sigma_{\rm rms}\simeq0.05$ K. For the $^{13}$CO data, a Gaussian with a width of 2.6 km s$^{-1}$ was applied to improve the noise level for the weaker emission features. The resultant noise level was $\sigma_{\rm rms}\simeq0.04$ K.

\bigskip

\subsection{{\rm Calibrations}}\label{sec:calib}

To confirm the accuracy and internal consistency of these data, we monitored the reference points in M17, W3, and OriKL every 2-3 hours in an observation day. The measured intensities of these sources are consistent within $\pm$7\% and the velocity accuracy is better than 0.3 km s$^{-1}$ for each span in which the receiver tuning was maintained (Yoda, 2008). However, the apparent intensities of the calibration sources changed slightly following each receiver tuning. This apparent change is likely due to differences in the image rejection ratio (IRR) of the receiver. In an ideal case, the IRR would be infinity and the receiver line intensity would be at a maximum. Therefore, we adjusted the intensity scale in each tuning span so that it would match with the maximum that had ever been measured. Since the IRR of both sidebands changes independently, we determined the scaling factor individually. After this correction, the main beam temperature at a reference point in OriKL, ($l$, $b$) = ($\timeform{209D.028765}$, $\timeform{-19D.375472}$) was 35.51$\pm$3.44 K in $^{12}$CO($J=2-1$) and 10.76$\pm$0.84 K in $^{13}$CO($J=2-1$). These values were derived using 5-point fitting around the peak position (Yoda, 2008) to reduce the pointing error effect.

The optical and radio pointing observations (see Nakajima et al., 2007) were made every 1-2 months, and we revised the pointing parameters so that the pointing accuracy would be better than $\timeform{1'}$. We checked this using the data of the reference points shown above, and confirmed that it would be about $\timeform{1'}$.

\bigskip

\section{Results}
\subsection{{\rm Overall View of Data}}\label{sec:res1}

\begin{figure}
\begin{center}
\FigureFile(150mm,200mm){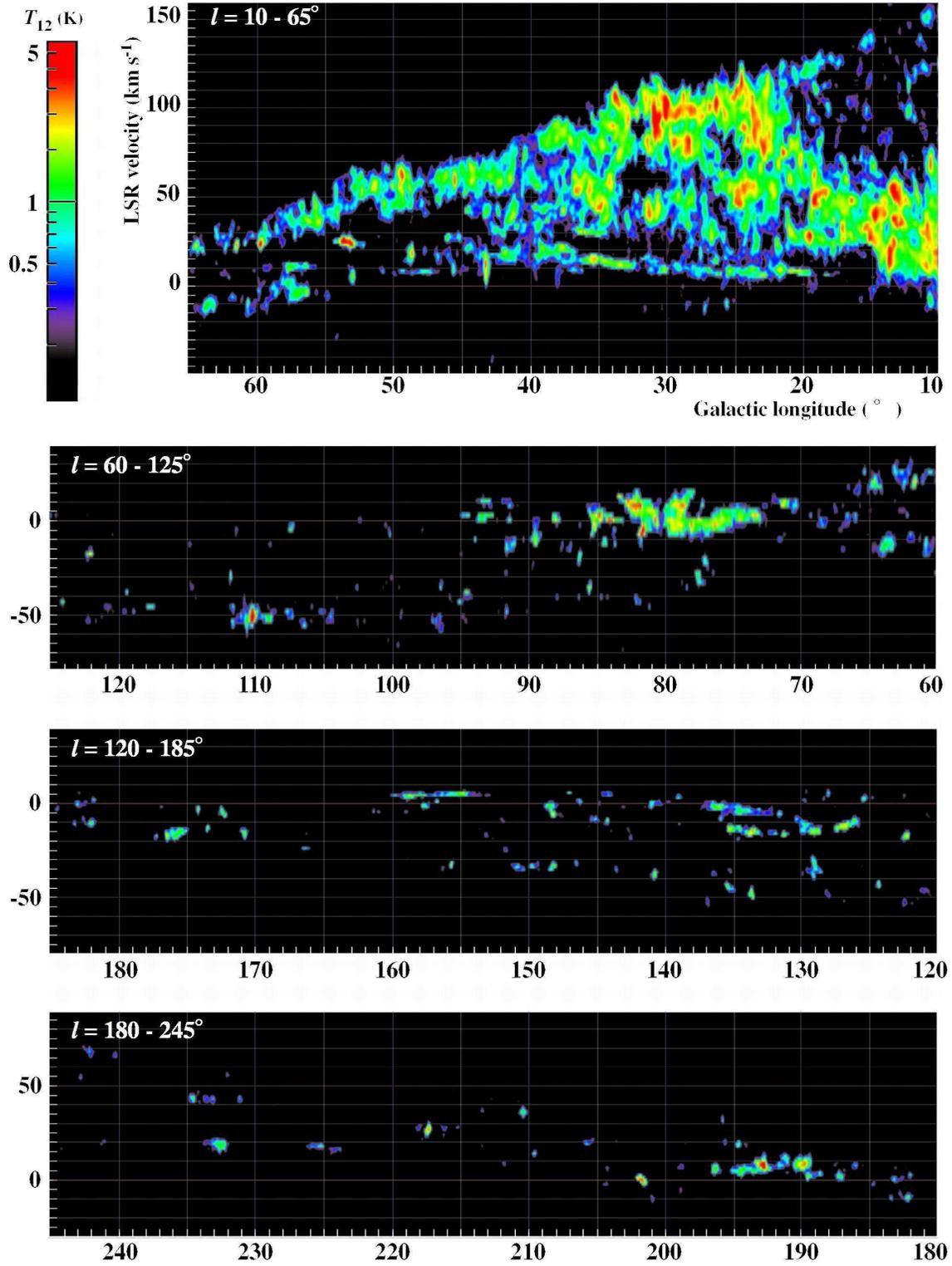}
\end{center}
\caption{The $l-v$ map of the $^{12}$CO($J=2-1$) line intensity. Each panel shows a different longitudinal range shown as the abscissa labels. The colors are shown in log-scale at the point where $T_{12}>3\sigma_{\rm rms}$.}
\label{fig:lv12co}
\end{figure}

\begin{figure}
\begin{center}
\FigureFile(150mm,200mm){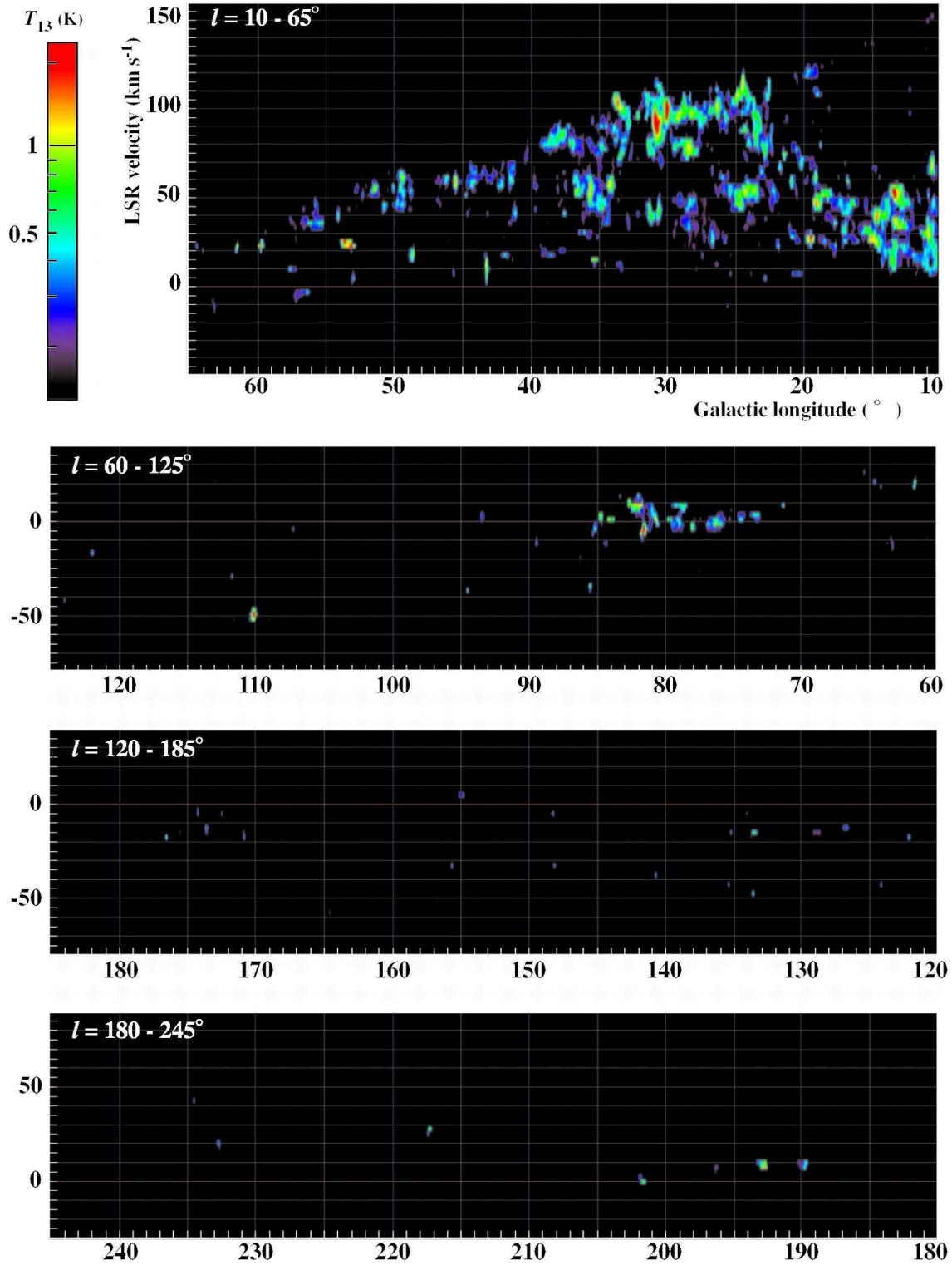}
\end{center}
\caption{The $l-v$ map of the $^{13}$CO($J=2-1$) line intensity. The data for $T_{13}$ are displayed in the same way as in Figure \ref{fig:lv12co}, though the color coding is different.}
\label{fig:lv13co}
\end{figure}

\begin{figure}
\begin{center}
\FigureFile(150mm,200mm){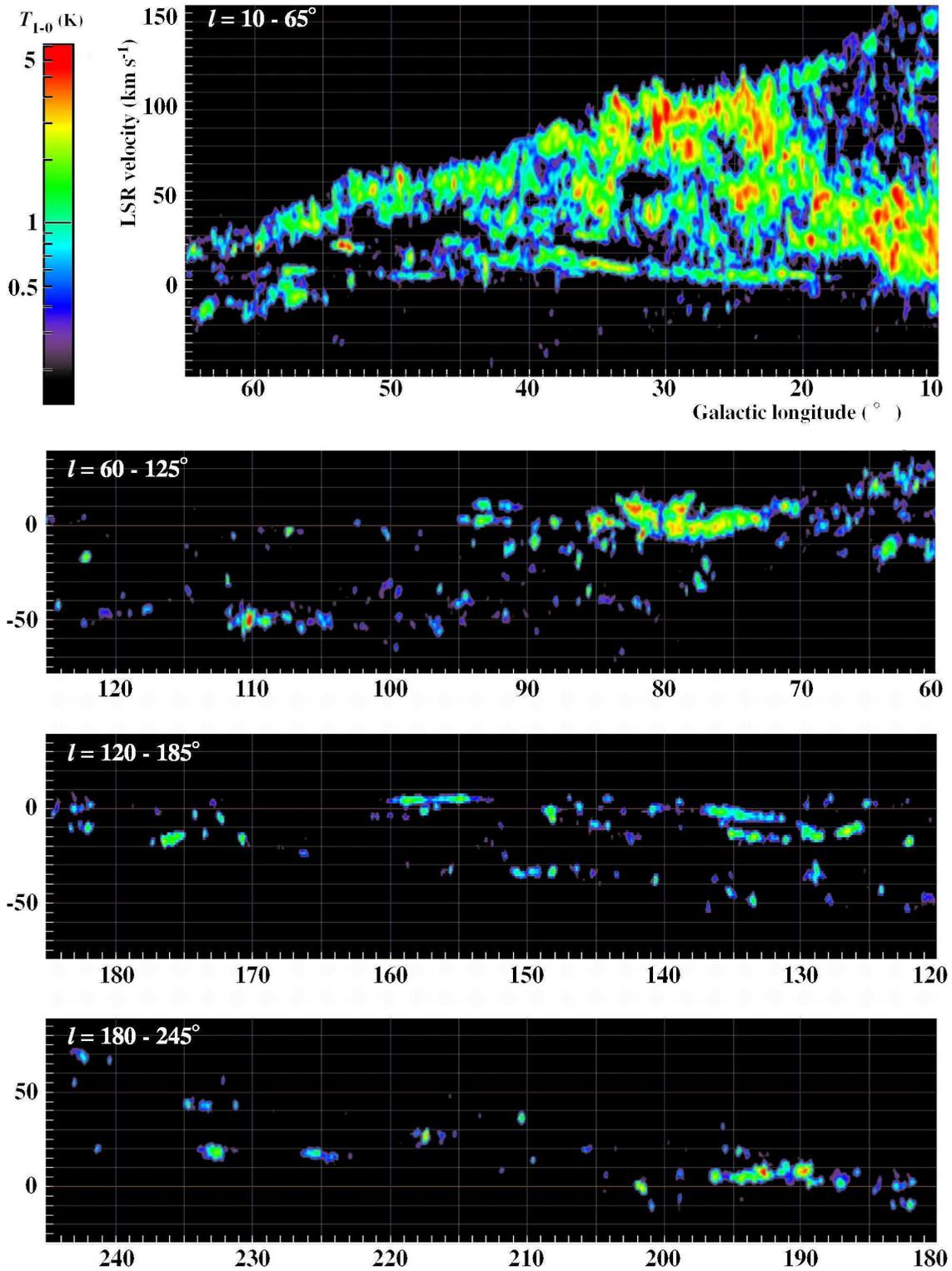}
\end{center}
\caption{The $l-v$ map of the $^{12}$CO($J=1-0$) line intensity produced from the observation data of Dame et al. (2001). The panels are displayed in the same way as in Figure \ref{fig:lv12co} and  \ref{fig:lv13co}, though the color coding is different.}
\label{fig:lvco10}
\end{figure}

Although the observed area has one pixel extent along the Galactic latitude, it had many pixels along the LSR velocity. Therefore, an overall view of our data can be shown in longitude-velocity ($l-v$) maps. Figure \ref{fig:lv12co} and Figure \ref{fig:lv13co} are $l-v$ maps of the main beam temperature in $^{12}$CO($J=2-1$), $T_{12}$, and in $^{13}$CO($J=2-1$), $T_{13}$, respectively. The velocity widths of a pixel on the $l-v$ plane are 1.3 km s$^{-1}$ in the $^{12}$CO line and 2.6 km s$^{-1}$ in the $^{13}$CO line. For comparison, we also show an $l-v$ map of the main beam temperature in $^{12}$CO($J=1-0$), $T_{1-0}$, which was constructed based on Dame et al. (2001) (Fig. \ref{fig:lvco10}). The velocity width is 1.3 km s$^{-1}$ and the grid spacing is 7.5 arcmin, which is twice as sparse as ours.

Since the $^{12}$CO($J=2-1$) and $^{13}$CO($J=2-1$) datasets were obtained simultaneously using the same receiver horn of the same telescope, there are no relative pointing errors, and the difference in beam size is less than 5\%. Moreover, our sampling and beam size coincide with those in the observations of Dame et al. (2001). We are therefore able to compare these data directly without any image processing.

\begin{figure}
\begin{center}
\FigureFile(150mm,200mm){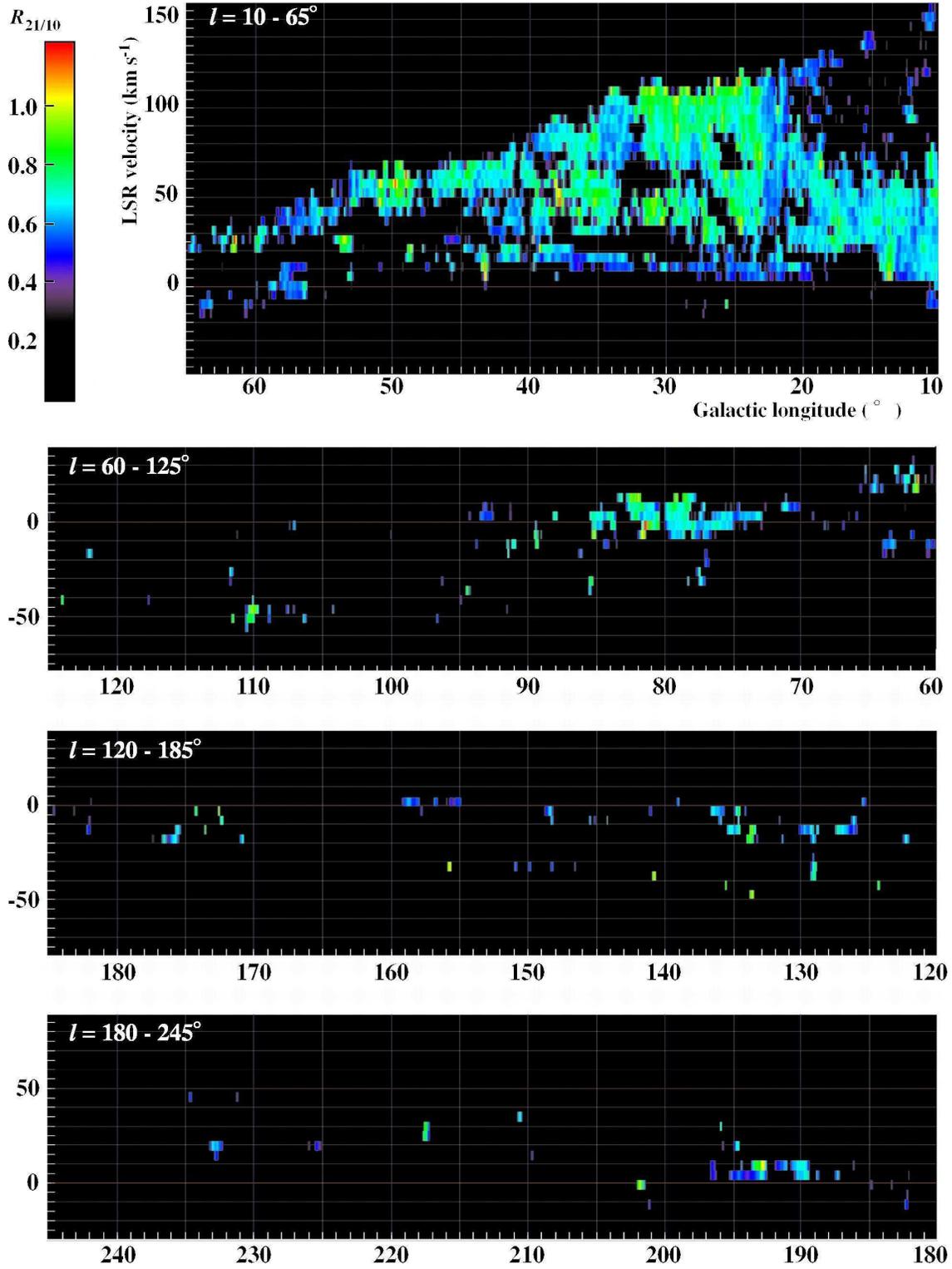}
\end{center}
\caption{The $l-v$ distribution of the ratio of $T_{12}$ to $T_{1-0}$. Both lines are averaged in a  5.2 km s$^{-1}$ velocity bin. The color coding is shown at the top-left. The data with $T_{12}\leq5\sigma_{{\rm rms}}\sim0.15$ K or $T_{1-0}\leq5\sigma_{{\rm rms}}\sim0.2$ K are shown in black.}
\label{fig:r2110}
\end{figure}

\begin{figure}
\begin{center}
\FigureFile(150mm,200mm){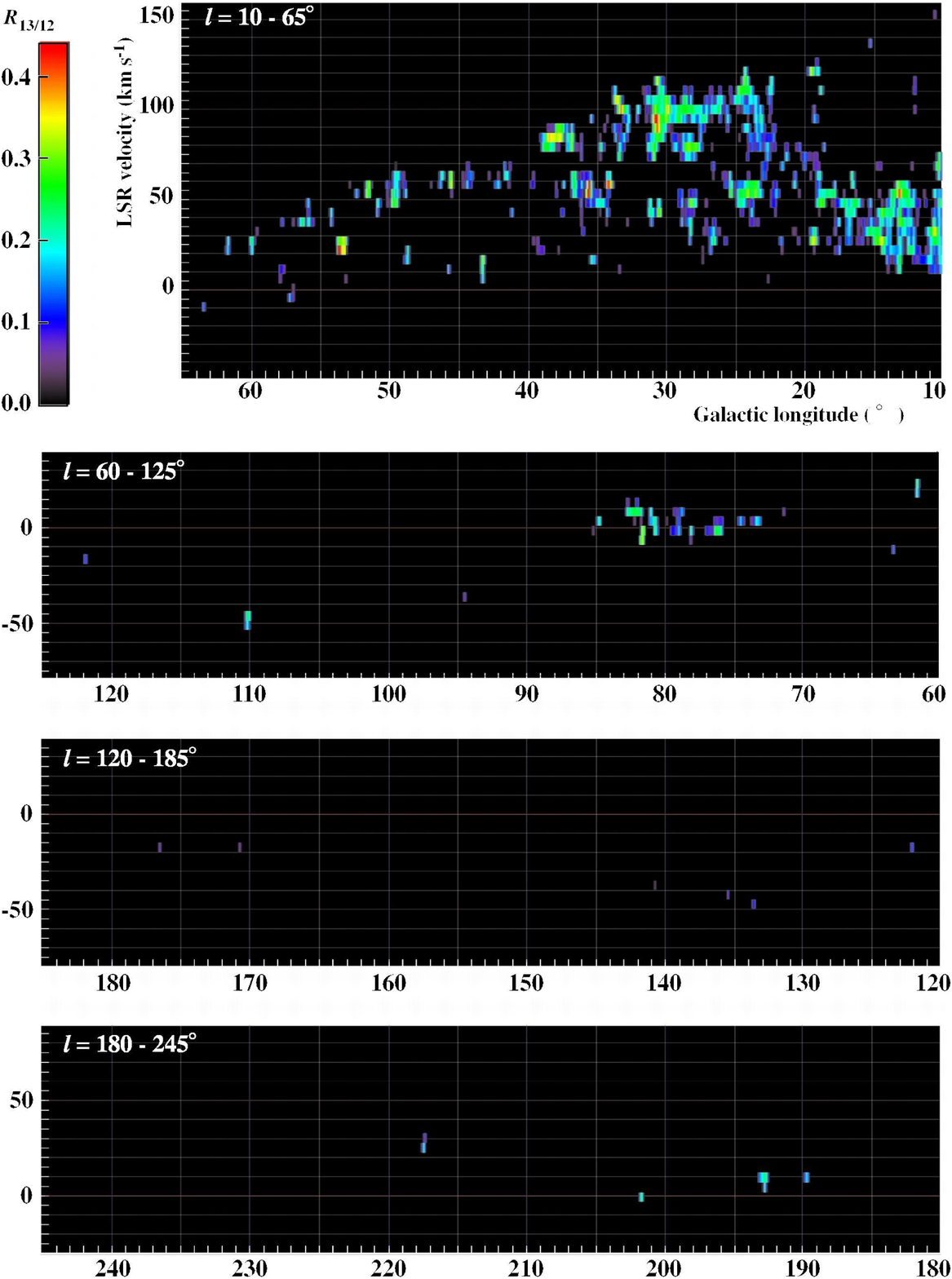}
\end{center}
\caption{The data of $T_{13}$ and $T_{12}$ are averaged in a 5.2 km s$^{-1}$ velocity bin. The ratio between them, $R_{13/12}$, is displayed on the $l-v$ plane. The data with $T_{13}\leq5\sigma_{{\rm rms}}\sim0.15$ K or $T_{12}\leq5\sigma_{{\rm rms}}\sim0.15$ K are shown in black.}
\label{fig:r1312}
\end{figure}

The $l-v$ maps of $T_{12}$, $T_{13}$ and $T_{1-0}$ look very similar. They are not, however, completely proportional to one another. For example, the $T_{12}$ intensity in the region of $l=10^\circ-20^\circ$ and $v_{\rm LSR}>60$ km s$^{-1}$ is much weaker than that of $T_{1-0}$, compared with its intensity in other regions. We put the $^{12}$CO($J=2-1$)/$^{12}$CO($J=1-0$) line intensity ratio, $R_{21/10}$, on the $l-v$ plane in Figure \ref{fig:r2110} and the $^{13}$CO($J=2-1$)/$^{12}$CO($J=2-1$) ratio, $R_{13/12}$, on the same plane in Figure \ref{fig:r1312}. The clipping levels are $>5\sigma_{\rm rms}$ for both lines. Both figures are drawn with a velocity width of 5.2 km s$^{-1}$ to reduce the noise level. This resolution is narrow enough to permit investigation of the Galactic structure because the typical velocity width of a cloud is similar to or wider than the resolution.

Figure \ref{fig:r2110} shows that most of the data pixels have a similar value. However, the standard deviation of $R_{21/10}$, 0.147, is larger than was expected from the system noise. $R_{21/10}$ actually changes from place to place and it shows a difference in the physical condition of the molecular gas. The average value of $R_{21/10}$ is 0.613. In limited regions around the terminal velocity, or 4 and 6 kpc from the Galactic center, for example ($l$, $v_{\rm LSR}$)$\simeq$($\timeform{50D}$, 60 km s$^{-1}$) and ($\timeform{25D}$, 110 km s$^{-1}$), some pixels show a remarkably high $R_{21/10}$ of about 1.0.

Figure \ref{fig:r1312} shows that the distribution of $R_{13/12}$ roughly corresponds to that of the line intensities of each line. For example, the peaks of $T_{13}$ at ($l$, $v_{\rm LSR}$)$\simeq$($\timeform{31D}$, 100 km s$^{-1}$) and ($\timeform{54D}$, 20 km s$^{-1}$) correspond to the peaks of $R_{13/12}$. There seems to be a general trend in which the more intense pixel shows the higher $R_{13/12}$. The average value of $R_{13/12}$ is 0.166.

\bigskip
\subsection{{\rm Correlations between Line Intensities}}\label{sec:res2}

\begin{figure}
\begin{center}
\FigureFile(75mm,75mm){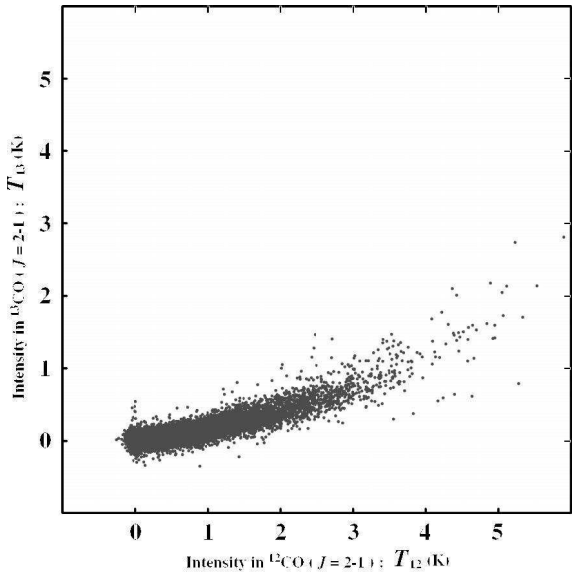}
\end{center}
\caption{The intensity correlation between $T_{13}$ and $T_{12}$. Each datum is averaged in a 5.2 km s$^{-1}$ velocity bin. The data plot forms a curve with increasing slope.}
\label{fig:ttplot1312}
\end{figure}

\begin{figure}
\begin{center}
\FigureFile(75mm,75mm){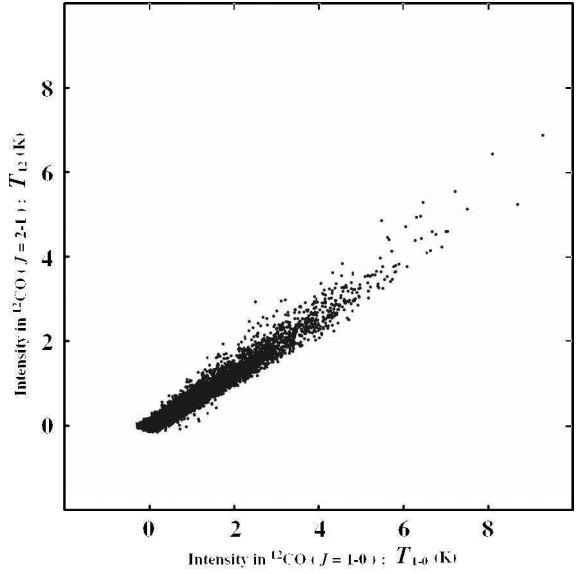}
\end{center}
\caption{The intensity correlation between $T_{1-0}$ and $T_{12}$. Each datum is averaged in a 5.2 km s$^{-1}$ velocity bin. The data plots forms a straight line.}
\label{fig:ttplot2110}
\end{figure}

In order to discover the characteristics of the Galactic disk gas, the correlations between the intensities in several lines are useful. We therefore constructed an intensity correlation plot between $T_{12}$ and $T_{13}$ in Figure \ref{fig:ttplot1312} and between $T_{1-0}$ and $T_{12}$ in Figure \ref{fig:ttplot2110}. In both figures, we only use the data in $l=10^\circ-65^\circ$  in order to reduce the effects of local molecular clouds and the Cygnus region. This approach is valid for investigating the global trend of the MWG, because this longitudinal range accounts for 84\% of the mass of the molecular gas in the entire observed area.

In the correlation between $T_{12}$ and $T_{13}$, most data points are aligned along a curve with increasing $R_{13/12}$ (Fig. \ref{fig:ttplot1312}). This means that $R_{13/12}$ is higher for the more intense data point, as we mentioned in the previous subsection. We will discuss the interpretation of this correlation curve in the next section.

In the correlation between $T_{12}$ and $T_{1-0}$, most data points are aligned in a linear correlation (Fig. \ref{fig:ttplot2110}). This means that most of the gas shows only a small variation in $R_{21/10}$. The slope of the regression line is 0.640. The observational error for the slope due to statistical noise, baseline error, and intensity fluctuation is estimated to be $\pm$0.051. The dispersion of all the measured data points gives $\pm$0.058 for the slope, which includes the intrinsic dispersion of $R_{21/10}$.

\begin{figure}
\begin{center}
\FigureFile(75mm,75mm){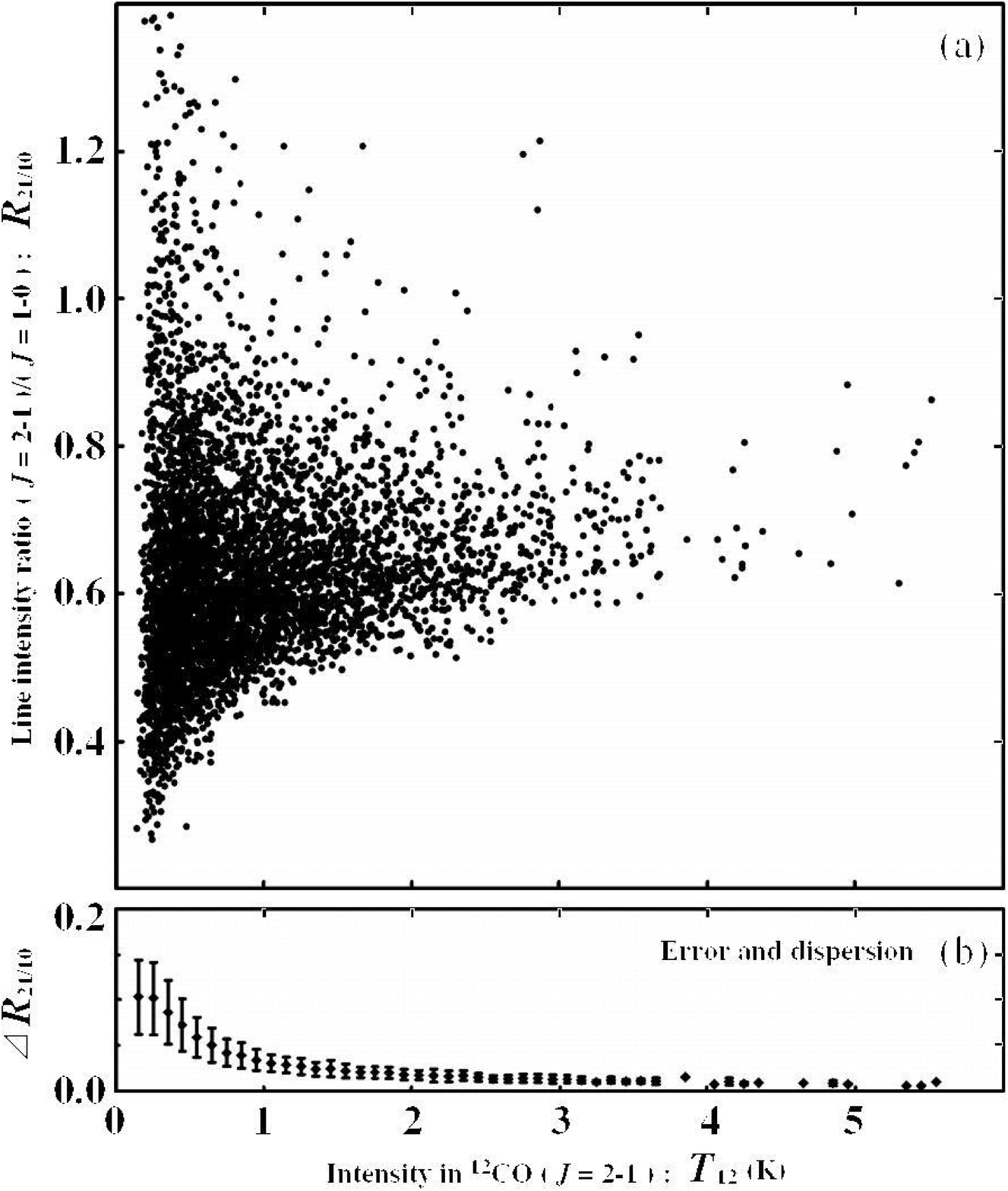}
\end{center}
\caption{
The upper panel (a) shows the correlation between $T_{12}$ and $R_{21/10}$. Each datum is averaged in a 5.2 km s$^{-1}$ velocity bin. The lower panel (b) shows the error of $R_{21/10}$, $\Delta R_{21/10}$. The $\Delta R_{21/10}$ depends on the value of $T_{12}$, $T_{1-0}$ and their noise levels through 
$\Delta R_{21/10} = \sqrt[]{\mathstrut \frac{1}{T_{1-0}}^2\Delta T_{12}^2+\frac{T_{12}}{T_{1-0}^2}^2\Delta T_{1-0}^2}$. 
The average value of $\Delta R_{21/10}$ in a $T_{12}$ bin is represented by a filled-in circle and its dispersion due to $T_{1-0}, \Delta T_{12}$ and $\Delta T_{1-0}$ are represented by bars.
}
\label{fig:lum_rat}
\end{figure}

In Figure \ref{fig:ttplot2110}, a number of data points are located above the regression line. This suggests that there may be a trend between line intensity and $R_{21/10}$. Next, we constructed a plot on the $T_{12}-R_{21/10}$ plane for the data on the $l-v$ map with $T_{12}\gtrsim0.25$ K (Figure \ref{fig:lum_rat}). Note that the error of the ordinate is not uniform along the abscissa, as shown in the lower panel of Fig. \ref{fig:lum_rat}. The error of the ratio is smaller than $\pm$0.05 for $T_{12}\gtrsim1$ K, while it reaches $\pm$0.1 for $T_{12}\lesssim0.5$ K. The error of $T_{12}$ is almost uniform at a typical value of $\sigma_{\rm rms}\simeq0.05$ K.

At $T_{12}\lesssim1$ K, the data points are scattered from $R_{21/10}\simeq$0.4 to 1.2. The scattered $R_{21/10}$ is real, because it is much larger than the observational error. In contrast, for $T_{12}\gtrsim3$ K, $R_{21/10}$ converges to about 0.6.

\bigskip
\section{Discussion}

We found that there are correlations between the line intensities in $^{12}$CO($J=2-1$) and $^{13}$CO($J=2-1$) and those in $^{12}$CO($J=2-1$) and $^{12}$CO($J=1-0$), although some data show deviations beyond the observational error. These relationships should reflect some constraint on the physical condition of the molecular gas in the Galactic disk. In this section, we discuss only the physical condition of the greater part of the gas that displays these relationships, or the global trend. We will not offer detailed discussion of the gas with a large deviation, which we will address in forthcoming papers.

To ascertain the actual relationship between the parameters of the physical condition of the molecular gas, we adopted a means of deriving them without the use of gas excitation models such as the large velocity gradient (LVG) model (Goldreich \& Kwan, 1974). We will give relationships empirically from the results presented in the previous section, based on the simple radiative transfer equation with the beam dilution effect:
\begin{equation}
T = \eta T_{\rm c} \bigl(1-\exp\left(-\tau\right)\bigr),
\label{eq:radtransf}
\end{equation}
where $\eta$ is the beam filling factor and $\tau$ is the optical depth in the line. Since our observation beam is larger than the typical size of a molecular cloud, the beam filling factor strongly affects the line intensity. $T_{\rm c}$ is the Rayleigh-Jeans line temperature\footnote{The Rayleigh-Jeans line temperature $T_{\rm c}$ is defined as
\begin{equation}
T_{\rm c} = \Biggl(\frac{h\nu}{k}\biggl(\exp\left(\frac{h\nu}{kT_{\rm ex}}\right)-1\biggr)^{-1}-T_{\rm bg}\Biggr),
\label{eq:tc}
\end{equation}
where $\nu$ and $T_{\rm ex}$ are the frequency and the excitation temperature of the transition, respectively. $T_{\rm bg}$ is the brightness temperature of the cosmic microwave background radiation (CMB) in the Rayleigh-Jeans approximation. Since the CMB is the 2.725 K blackbody, $T_{\rm bg,12}=0.19$ K, $T_{\rm bg,13}=0.22$ K and $T_{\rm bg,1-0}=0.84$ K at the corresponding frequencies. $T_{\rm c}$ is close to $T_{\rm ex}$ in the case that $T_{\rm ex}\gg \frac{h\nu}{k}$ and $T_{\rm ex}\gg T_{\rm bg}$.} that is introduced to simplify the expression.

In subsection \ref{sec:rep1312} and \ref{sec:rep2110}, we derive the empirical relationships between the parameters of the molecular gas. Based on these, we discuss the typical physical condition of the gas in subsection \ref{sec:T_and_n}. In subsection \ref{sec:unusual}, we briefly discuss the data points that deviated from the typical relationship presented in the previous section. In this paper, we assume that the emitting region of a beam is uniform.

\bigskip
\subsection{{\rm Correlation between }$T_{12}${\rm and }$T_{13}${\rm and Two Empirical Relationships of the CO gas}}\label{sec:rep1312}

From the correlation curve between the line intensities in $^{12}$CO($J=2-1$) and $^{13}$CO($J=2-1$), we can obtain two restrictions on these lines. One is that the optical depth in the $^{12}$CO line is much larger than that in the $^{13}$CO line. If it is not, then the correlation curve must be close to a straight line. This is consistent with the relative abundance of these two molecular species; it is reported to be about 60 in the Galactic disk (e.g., about 59 by Liszt \& Lucas, 1998; about 76 by Stahl et al., 2008). With this abundance, the optical depth in the $^{12}$CO line is expected to be several tens of times as thick as that in the $^{13}$CO line.

This means that the $^{12}$CO lines should be thick in most cases. Although many observations have already revealed that the $^{12}$CO line is optically thick in the nearby molecular clouds, our findings show that it is also thick in most of the molecular gas beyond the boundaries of mapped molecular clouds.

In this case, we can obtain the following relationships, using equation \ref{eq:radtransf}:
\begin{eqnarray}
 T_{13} &=& \eta_{13} T_{\rm c,13}\bigl(1-\exp\left(-\tau_{13}\right)\bigr),
\label{eq:t13}
\\
 T_{12} &=& \eta_{12} T_{\rm c,12}.
\label{eq:t12}
\end{eqnarray}
The approximation in the second equation is valid within 5\%, if $\tau_{12}>3$.

We should note that some data should be optically thin in $^{12}$CO($J=2-1$). In particular, the data points with $R_{21/10}\gtrsim1.0$ are difficult to explain if the line is opticaly thick. Such data points, however, are less than 7\% and such gas is beyond the scope of the present discussion.

Equations \ref{eq:t13} and \ref{eq:t12} have five parameters, but the effective number of free parameters in only three, because both $\eta_{13}T_{\rm c,13}$ and $\eta_{12}T_{\rm c,12}$ affect the equations as single parameters. Therefore, the correlation curve between the $^{12}$CO and the $^{13}$CO lines gives a number of relationships among $\eta_{13}T_{\rm c,13}, \eta_{12}T_{\rm c,12}$ and $\tau_{13}$.

Within the first order approximation, they must be
\begin{eqnarray}
\frac{\eta_{13}T_{\rm c,13}}{\eta_{12}T_{\rm c,12}} &=& \alpha,
\label{eq:modeleq1}
\\
\eta_{13}T_{\rm c,13} &=& \beta\cdot\tau_{13},
\label{eq:modeleq2}
\end{eqnarray}
where $\alpha$ and $\beta$ are constants. To estimate these two cnstants, we evaluate $\chi^2$ as defined by
\begin{equation}
\chi^2=\sum \Bigl(T^{\rm obs}_{13}-T^{\rm exp}_{13}\Bigr)^2+\sum\Bigl(T^{\rm obs}_{12}-T^{\rm exp}_{12}\Bigr)^2,
\label{eq:defkaisq}
\end{equation}
where the superscript ``obs'' represents the observed value, and ``exp'' represents the value expected from the observed value using equation \ref{eq:modeleq1} and \ref{eq:modeleq2}. The calculation is carried out for the averaged $T^{\rm obs}_{13}$ over each $T^{\rm obs}_{12}$ bin with a 0.1 K step in order to avoid over-weighting the data points with small $T^{\rm obs}_{12}$ and to obtain a globally fitted curve.

\begin{figure}
\begin{center}
\FigureFile(100mm,67mm){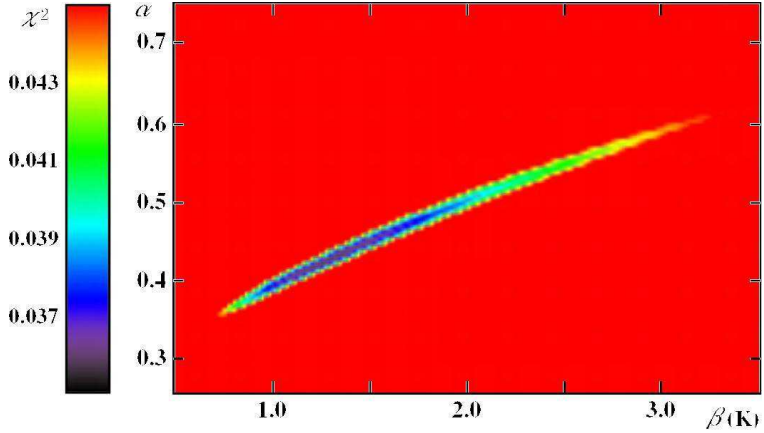}
\end{center}
\caption{The distribution of $\chi^2$ given by equation \ref{eq:defkaisq} on the $\alpha-\beta$ plane, which are the two parameters of the model expressed in equations \ref{eq:modeleq1} and \ref{eq:modeleq2}. The $\chi^2$ value is low along a line and takes its minimum at ($\alpha$,$\beta$) = (0.4, 1.3K).}
\label{fig:kaisqmap}
\end{figure}

\begin{figure}
\begin{center}
\FigureFile(75mm,75mm){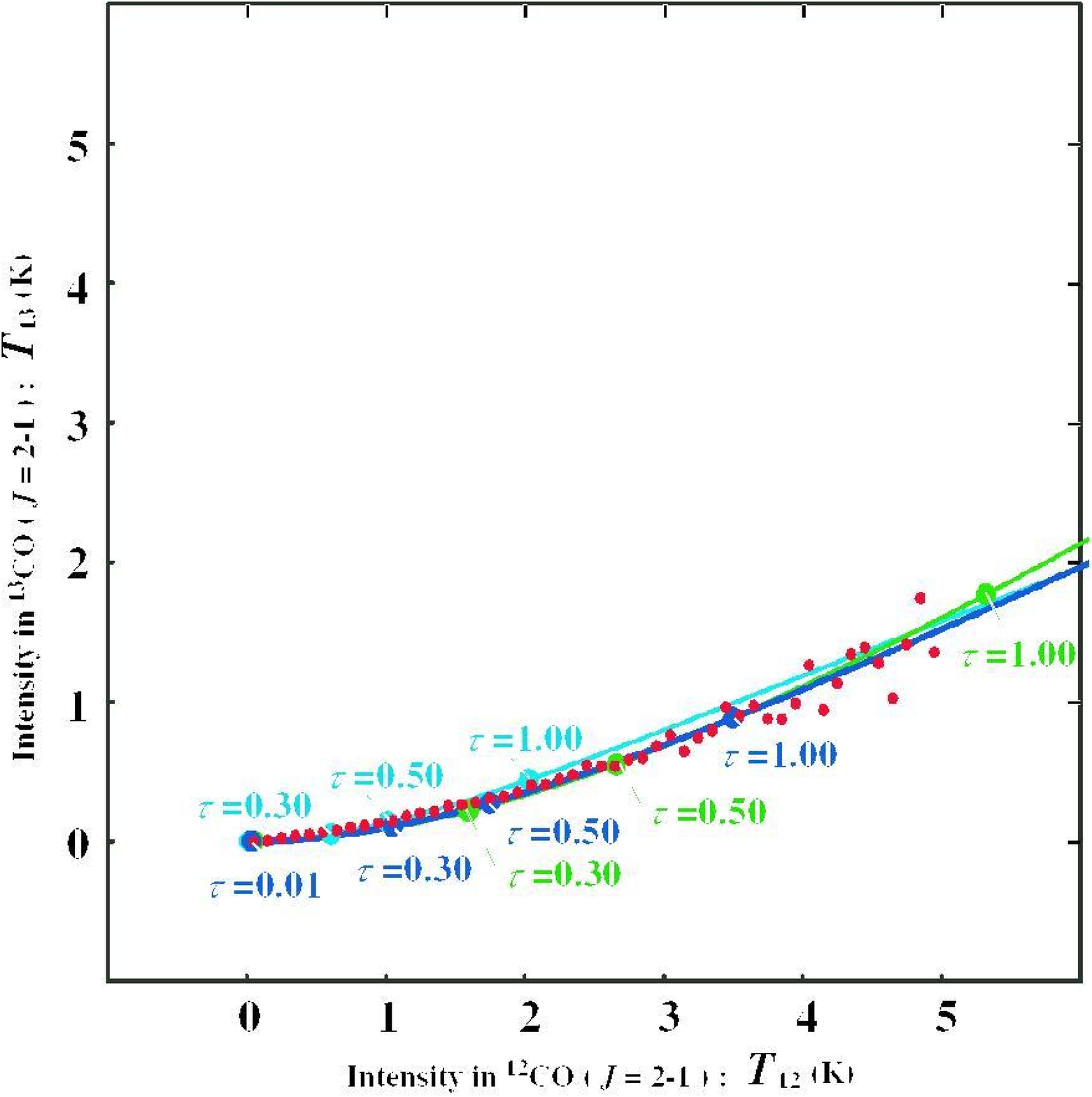}
\end{center}
\caption{The correlation curve produced by our model equations \ref{eq:modeleq1} and \ref{eq:modeleq2}. The blue, cyan and green solid lines are correspond to ($\alpha$, $\beta$) =  (0.4, 1.3 K), (0.3, 0.7 K), and (0.5, 2.8 K), respectively. The red dots are the $T_{12}-T_{13}$ intensity correlation used in the $\chi^2$ fitting. The filled-in circles in each color correspond to $\tau_{13} = $0.01, 0.30, 0.50, and 1.00, respectively. }
\label{fig:repcurve}
\end{figure}

The resulting distribution of $\chi^2$ on the $\alpha-\beta$ plane is shown in Figure \ref{fig:kaisqmap}. It indicates that the $\chi^2$ value is small along a line. Although ($\alpha$, $\beta$) = (0.4, 1.3 K) gives the smallest $\chi^2$, the correlation curve is also produced in the case of (0.3, 0.7 K) or (0.5, 2.8 K) (see Fig. \ref{fig:repcurve}).

\bigskip

What is the astrophysical meaning of these two constants? Equations \ref{eq:modeleq1} and \ref{eq:modeleq2} would represent the ``common sense'' of the molecular gas in the Galactic disk. In the following discussion, we elaborate this idea using the terminology of astrophysics.

The constant $\alpha$ is the ratio of $\eta_{13}T_{\rm c,13}$ and $\eta_{12}T_{\rm c,12}$. We divide it as a product of $\eta_{13}/\eta_{12}$ and $T_{\rm c,13}/T_{\rm c,12}$.

$\eta_{13}/\eta_{12}$ is the ratio of the projected sizes of the emitting region. As shown in the first part of this subsection, the $^{12}$CO line is optically much thicker than the $^{13}$CO line. This suggests that the emitting region in the $^{12}$CO line should be larger than that in the $^{13}$CO line. Sawada et al. (2001) estimated that in the Galactic center region, $\eta_{13}/\eta_{12}$ is about 2/3 or 0.7. Although the actual value of the $\eta_{13}/\eta_{12}$ in the Galactic disk should be different from that in the Galactic center region, it is difficult to estimate. For now, then, we therefore assume that it is about 0.7.

In the case that $\eta_{13}/\eta_{12}$ is constant over the Galactic disk, the constant $\alpha$ means that $T_{\rm c,13}/T_{\rm c,12}$ is constant. This does not directly indicate that $T_{\rm ex,13}/T_{\rm ex,12}$, because $T_{\rm c}$ is not linear to $T_{\rm ex}$. For $\eta_{13}/\eta_{12}=0.7$, $\alpha=0.4\pm0.1$ gives $T_{\rm c,13}/T_{\rm c,12}=0.6\pm0.1$, which corresponds to $T_{\rm ex,13}/T_{\rm ex,12}\simeq0.7\pm0.1$.

Since the $^{12}$CO line can be emitted from a less dense gas than the $^{13}$CO line can, $\eta_{12}$ should be larger than $\eta_{13}$ for a cloud with an envelope of less dense gas, and $\eta_{12}$ should be equal to $\eta_{13}$ for a cloud without the envelope. Therefore, the upper limit of $\eta_{13}/\eta_{12}$ should be unity. In this case, $T_{\rm ex,13}/T_{\rm ex,12}$ is about 0.5. The lower limit of $\eta_{13}/\eta_{12}$ is difficult to estimate without a model of the density structure of a cloud. If we presume $\eta_{13}/\eta_{12}=0.5$, then $T_{\rm ex,13}/T_{\rm ex,12}$ is about 0.8.

The other constant $\beta$ should show the condition of the $^{13}$CO emission, because it is the linear coefficient of $\tau_{13}$ to $\eta_{13}T_{\rm c,13}$. The derived value of $\beta$ changes from 0.7 K to 2.8 K depending on the value of $\alpha$. Equation \ref{eq:modeleq2}, which defines $\beta$, is qualitatively reasonable once we consider the following discussion. A larger gas cloud gives a larger $\tau$ and a larger $\eta$. In a gas cloud of fixed size in which the gas is subthermally excited and in which photon trapping is essential for molecule excitation, a denser gas gives a larger $\tau$, and a larger $\tau$ gives a higher $T_{\rm ex}$ and $T_{\rm c}$ for the cloud. Although it is difficult to show which is more affected by $\tau_{13}$, $\eta_{13}$ or $T_{\rm c,13}$, equation \ref{eq:modeleq2} may indicate a relationship between $\tau_{13}$ and $\eta_{13}$, because both parameters are directly affected by cloud size.

\bigskip

\subsection{{\rm Ratio of }$T_{12}${\rm and }$T_{1-0}${\rm and Excitation of }$^{12}${\rm CO}}\label{sec:rep2110}

From the linear correlation between the line intensities in $^{12}$CO($J=2-1$) and $^{12}$CO($J=1-0$), we can obtain a restriction on $^{12}$CO excitation. We showed $R_{21/10}=0.640\pm0.058$ in Section \ref{sec:res1}. Since the $^{12}$CO lines are optically thick, $R_{21/10}$ is expressed as
\begin{equation}
R_{21/10} = \frac{\eta_{12}T_{\rm c,12}}{\eta_{1-0}T_{\rm c,1-0}}.
\label{eq:r2110eta}
\end{equation}
Although we cannot evaluate the exact value of the beam filling factors, $\eta_{12}$ should be close to $\eta_{1-0}$, because the excitation conditions for both emission lines are similar and the apparent size in both lines are close for a model cloud (Sakamoto et al., 1997). In this case, we can deduce
\begin{equation}
R_{21/10} = \frac{T_{\rm c,12}}{T_{\rm c,1-0}}.
\label{eq:r2110}
\end{equation}
Therefore, $R_{21/10}=0.640\pm0.058$ indicates that $^{12}$CO($J=2-1$) is subthermally excited.

In the previous subsection, we concluded that $^{12}$CO($J=2-1$) is optically thick. This large optical depth allows for the small probability that the emission line photons may escape and make the gas thermalized. However, moderate opacity can give both subthermal excitation and optically thick approximation for the line intensities. For example, the gas with $T_{\rm K}\lesssim20$ K, $n$(H$_2$)$=10^{2.0-2.5}$cm$^{-3}$ and $N$($^{12}$CO)/d$v=10^{16-17}$ cm$^{-2} $ km$^{-1}$ s shows subthermality and $\tau_{12}\simeq5$, using the LVG model.

\begin{figure}
\begin{center}
\FigureFile(100mm,60mm){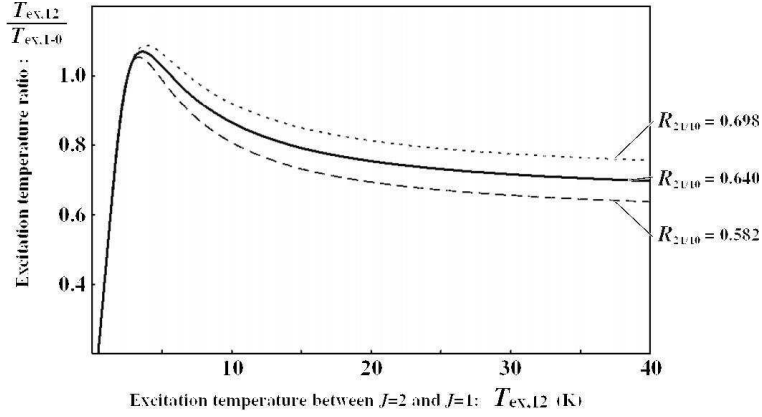}
\end{center}
\caption{The theoretical loci of $R_{21/10}=\frac{T_{\rm c,12}}{T_{\rm c,1-0}}=0.640$ (solid line), 0.640-0.058 (dahsed line), and 0.640+0.058 (dotted line) on the $T_{\rm ex,12}-\frac{T_{\rm ex,12}}{T_{\rm ex,1-0}}$ plane, as calculated with equation \ref{eq:tc}. If $T_{\rm ex,12}$ changes between 13 K and 40 K, the constant $T_{\rm ex,12}/T_{\rm ex,1-0}$ gives the same $R_{21/10}$ within the observational error. If $T_{\rm ex,12}$ is colder than 13 K, the constant $R_{21/10}$ requires a higher $T_{\rm ex,12}/T_{\rm ex,1-0}$ for the colder gas.}
\label{fig:tex2110}
\end{figure}

A constant $T_{\rm c,12}/T_{\rm c,1-0}$ does not always give a constant $T_{\rm ex,12}/T_{\rm ex,1-0}$, because $T_{\rm c}$ is linear to $T_{\rm ex}$ only when $T_{\rm ex}\gg h\nu/k$ and $T_{\rm ex}\gg T_{\rm bg}$. Figure \ref{fig:tex2110} shows the loci of $T_{\rm c,12}/T_{\rm c,1-0}=0.640\pm0.058$ on the $T_{\rm ex,12}-T_{\rm ex,12}/T_{\rm ex,1-0}$ plane. The value of $T_{\rm ex,12}/T_{\rm ex,1-0}$ increases near $T_{\rm ex,12}=T_{\rm bg}$. From our data, we found that most of the molecular gas shows almost constant $R_{21/10}$. If $T_{\rm ex,12}$ is not fairly uniform over the Galactic disk, Fig. \ref{fig:tex2110} suggests that there is a lower limit on it, because the gas that is colder than this temperature must have a higher $T_{\rm ex,12}/T_{\rm ex,1-0}$ than the warmer gas in order to keep the same $R_{21/10}$ value. There is a non-negligible amount of molecular gas that is warmer than 20 K. Therefore, we can conclude that $T_{\rm ex,12}$ is warmer than 13 K. In this case, $T_{\rm ex,12}/T_{\rm ex,1-0}$ is 0.7 and $T_{\rm ex,1-0}$ is higher than about 19 K.

Combined with $T_{\rm ex,13}/T_{\rm ex,12}$ given in the previous subsection, we can estimate the lower limit of $T_{\rm ex,13}$. For $T_{\rm ex,13}/T_{\rm ex,12}\simeq0.7$, it is about 9 K. Although we did not make observations in $^{13}$CO($J=1-0$) with the same beam size, its excitation temperature should be higher than 12 K, if the ratio of the excitation temperature of $^{13}$CO($J=2-1$) and $^{13}$CO($J=1-0$) is similar to $T_{\rm ex,12}/T_{\rm ex,1-0}$, $\simeq0.75$. Using higher resolution (44 arcsec) data for the molecular clouds in the 4kpc-ring region, Rathborne et al. (2009) estimated that the excitation temperature in $^{13}$CO($J=1-0$) line is 9 K. Our estimation is not inconsistent with theirs, because these estimations may have large ambiguity due to the many assumptions involved.

\begin{figure}
\begin{center}
\FigureFile(75mm,75mm){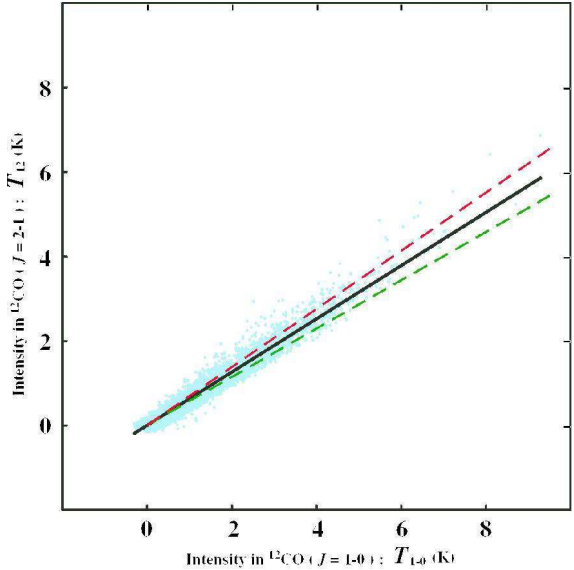}
\end{center}
\caption{Regression line derived by the $\chi^2$ fitting, which has a slope of 0.640, and lines with a slope of 0.640$\pm$0.058 (dashed lines). The background is the $T_{1-0}-T_{12}$ intensity correlation plot shown in Figure \ref{fig:ttplot2110}.}
\label{fig:rep2110}
\end{figure}

\bigskip

\subsection{{\rm Gas Kinematic Temperature and Typical Gas Density.}}\label{sec:T_and_n}

As shown in the previous section, the lower limit of $T_{\rm ex,1-0}$ is about 19 K. This means that the lower limit of the gas kinematic temperature $T_{\rm K}$ should also be about 19 K. We believe that $T_{\rm K}$ is not much higher than $T_{\rm ex,1-0}$, since $^{12}$CO($J=1-0$) is optically thick and the energy level of $J=1$ is only 5.5 K greater than the ground level.

It is very difficult to estimate the gas density in molecular clouds from low resolution data that is insufficient to resolve individual clouds. We can, however, discuss the typical gas density in a cloud with the help of the derived values given in previous subsections.

With low resolution data, the beam filling factor is of essential importance in deriving the physical property of interstellar gas. In the optically thick case, it was given as the ratio of the measured brightness temperature to the Rayleigh-Jeans line temperature, as shown in equation \ref{eq:t12}. Since all of data points show $T_{12}<7$ K and about 80\% of the data points in our sample show $T_{12}<1.5$ K, $\eta_{12}$ must be less than 0.7 and should be less than 0.14 for most of the data points, though some of them may be different from those for gas with a homogeneous property, which is discussed in this paper.

We can estimate the volume filling factor $\zeta$ which is the volume fraction of line emitting region to the sampling volume in an observation beam, from the beam filling factor $\eta$, when the distribution of the emitting region is isotropic in the sampling volume. We can neglect the overlapping of clouds in the sky because the actual beam filling factor is much smaller than unity. Thus we obtain
\begin{equation}
\zeta \simeq \frac{r}{s}\eta^{\frac{3}{2}},
\label{eq:vfil}
\end{equation}
where $r$ is the spatial resolution at the subcentral point, or at the average distance to the two positions corresponding to the same velocity, given by the beam size, and $s$ is the path length of a data cell. Using equation \ref{eq:t12}, we estimated the upper limit of $\eta_{12}$ for each data point. Using a kinematic model of the MWG, we can also estimate $r$ and $s$ for each data point.

\begin{figure}
\begin{center}
\FigureFile(100mm,67mm){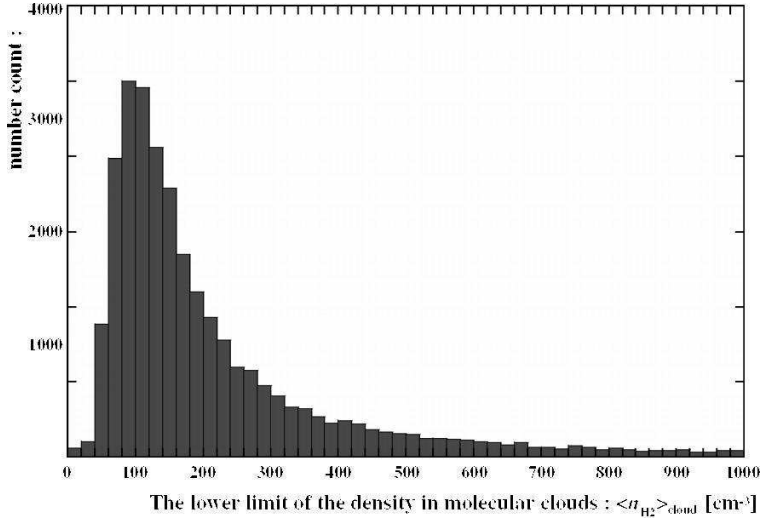}
\end{center}
\caption{Histogram of the lower limit of the gas density in molecular clouds $\langle n$(H$_2$)$\rangle_{\rm cloud}$ as calculated by $\langle n$(H$_2$)$\rangle_{\rm beam}/\zeta$. The histogram shows a peak at $\langle n$(H$_2$)$\rangle_{\rm cloud}\simeq$100cm$^{-3}$. }
\label{fig:cdensity}
\end{figure}

Many investigations of the Galactic plane and external galaxies use the assumption that $T_{1-0}$ is proportional to the H$_2$ column density $N$(H$_2$) through the CO-to-H$_2$ conversion factor $X$ for the first order approximation. Using the path length $s$, we can estimate the H$_2$ volume density averaged over the beam and the velocity bin as
\begin{equation}
\langle n({\rm H}_2)\rangle_{\rm beam}=\frac{X}{s}T_{1-0}.
\label{eq:bdensity}
\end{equation}
According to the definition of the volume filling factor, $\langle n$(H$_2$)$\rangle_{\rm beam}$ is equal to the gas density in molecular clouds $\langle n$(H$_2$)$\rangle_{\rm cloud}$ diluted by $\zeta$. Therefore, we can estimate the lower limit of $\langle n$(H$_2$)$\rangle_{\rm cloud}$ for each data point. A histogram of this shown in Figure \ref{fig:cdensity} with $X=1.8\times10^{20}$ cm$^{-2}$ (K km s$^{-1}$)$^{-1}$ (Dame et al., 2001). The estimated lower limit of $\langle n$(H$_2$)$\rangle_{\rm cloud}$ is typically more than 100 cm$^{-3}$, which is consistent with the results derived from the data of a nearby molecular cloud (e.g., Sakamoto et al., 1994) or molecular gas in a Galactic disk of a galaxy using a CO excitation model (e.g., Muraoka et al., 2007).

\bigskip

\subsection{{\rm Data Points that Deviate from Main Trend}}\label{sec:unusual}

Until the previous subsection, our discussion focused on the global trend of the data. A number of data points in Figure \ref{fig:ttplot1312}-\ref{fig:lum_rat}, however, deviate from the trend. In this subsection, we make observations on the characteristics of these data points.

In subsection \ref{sec:res2}, we showed that $R_{21/10}$ has a large intrinsic dispersion at a weaker $T_{12}$ (Fig. \ref{fig:lum_rat}). We believe that such data points with weak $T_{12}$ are located in the inter-arm region, because the spiral arm is traced by concentrations of molecular gas, which indicates intense $T_{12}$. In fact, observations of external galaxies have shown that intense features in a CO line are distributed on the spiral arm (e.g., Kuno et al., 2007). The data points with intense $T_{12}$ are continuous in the $l-v$ plane and they correspond to spiral arm
(Fig. \ref{fig:lv12co}). It is certain that $R_{21/10}$ is controlled by the temperature and/or density, though it is difficult to factorize their respective contributions. Therefore, the large dispersion at $T_{12}$ suggests that the physical condition is more heterogeneous in the inter-arm region than in the spiral arm.

In Figure \ref{fig:ttplot2110}, a number of plots deviated from the regression line toward the higher $T_{12}$. These data points are also located far from the majority in the $T_{12}-R_{21/10}$ plane (Fig. \ref{fig:lum_rat}). One would not expect the data points in an area of $T_{12}\simeq1-4$ K and $R_{21/10}\gtrsim0.86$ to be optically thick, because $R_{21/10}\gtrsim0.86$ requires $T_{\rm ex,12}>T_{\rm ex,1-0}$ in the optically thick case. On the $l-v$ plane, most of them are located near H$\emissiontype{II}$ regions either at the terminal velocity in $l\simeq25^\circ-30^\circ$ and $l\simeq50^\circ$ or near W49 and W44.

The $T_{12}-T_{13}$ correlation plot (Fig. \ref{fig:ttplot1312}) also has a number of data points that are located far from the correlation curve. Some plots at $T_{12}\simeq5$ K showed remarkably lower $T_{13}$ than the global trend. These data points show a fairly higher $R_{21/10}$ than the averaged value and should have lower opacity in $^{12}$CO($J=2-1$). They correspond to a molecular cloud at $l\simeq13.5^\circ$ and $v_{\rm LSR}\simeq10$ km s$^{-1}$, which is overlaid on a low mass X-ray binary, GX13+1 (White et al., 1978). This cloud would exist under the unique physical condition. High resolution observations would be helpful to resolve this issue.

\bigskip

\section{Summary}

In order to investigate the global physical condition of the MWG, we carried out simultaneous observations of  $^{12}$CO and $^{13}$CO $J=2-1$. We covered the Galactic plane at  $b=0^\circ$ over $l=10^\circ-245^\circ$ with a $\timeform{3'.75}$ grid using the AMANOGAWA telescope. Using the $^{12}$CO($J=1-0$) dataset of Dame et al. (2001), which has almost the same spatial resolution, we showed the distribution of the $^{12}$CO($J=2-1$)/$^{12}$CO($J=1-0$) and the $^{13}$CO($J=2-1$)/$^{12}$CO ($J=2-1$) ratios on the $l-v$ plane and the intensity correlations between their intensities.

We found that for the majority of data points, the intensity correlation between $^{12}$CO($J=2-1$) and $^{13}$CO($J=2-1$) is aligned along a curve, and that between $^{12}$CO($J=2-1$) and $^{12}$CO($J=1-0$) lies along a straight line with an intensity ratio of $R_{21/10}=0.640$. From these results, we conclude that the optical depth in the $^{12}$CO($J=2-1$) line is much greater than that in $^{13}$CO($J=2-1$), and most of the data points follow the two simple relationships shown in equations \ref{eq:modeleq1} and \ref{eq:modeleq2}. Using the best fit values of parameters $\alpha$ and $\beta$ in these equations, we estimated typical values for the physical condition of the molecular gas in the Galactic disk as follows: the excitation temperatures in $^{12}$CO($J=1-0$), $^{12}$CO($J=2-1$), and $^{13}$CO($J=2-1$) are higher than 19 K, 13 K, and 9 K, respectively. The gas kinetic temperature is higher than 19 K. The H$_2$ volume density in a cloud is more than 100 cm$^{-3}$. The beam filling factor is much less than 0.7.

We found a number of data points that deviate from the global trend. This means that the gas temperature and density in these locations are different from the typical values we obtained. Some of these are located close to H$\emissiontype{II}$ regions or an X-ray source, while the others are located in the interarm region.

\bigskip

\bigskip

The AMANOGAWA telescope is a joint project of the University of Tokyo, the National Astronomical Observatory of Japan, Osaka Prefecture University, and Tokyo Gakugei University. We especially thank the staff at the Nobeyama Radio Observatory, Nobeyama Solar Radio Observatory, and the Advanced Technology Center for their support in operating the telescope. We would also like to thank the builders of the original telescope system, before the upgrade. Finally, we thank our referee for the valuable advice given to improve this paper. This study was financially supported by Japan Society for Promotion of Science as Grants-in-Aid for Scientific Research C(18540232). T.Y. was financially supported by Global COE Program ``the Physical Science Frontier'', MEXT, Japan.


\end{document}